# Applications of Multivariate Statistical Methods and Simulation Libraries to Analysis of Electron Backscatter Diffraction and Transmission Kikuchi Diffraction Datasets

Angus J Wilkinson[1], David M Collins[1,2], Yevhen Zayachuk[1], Rajesh Korla[1,3], Arantxa Vilalta-Clemente[1,4]

[1]   Department of Materials, University of Oxford, Parks Road, Oxford, OX1 3PH, UK.
[2]   now at School of Metallurgy and Materials, University of Birmingham, Edgbaston, B15 2TT, UK.
[3]   now at Department of Materials Science & Metallurgical Engineering, Indian Institute of Technology, Hyderabad, India
[4]   now at Groupe de Physique des Matériaux, Université de Normandie Rouen, INSARouen, CNRS 6634, 76000 Rouen, France

## Abstract

Multivariate statistical methods are widely used throughout the sciences, including microscopy, however, their utilisation for analysis of electron backscatter diffraction (EBSD) data has not been adequately explored.  The basic aim of most EBSD analysis is to segment the spatial domain to reveal and quantify the microstructure, and links this to knowledge of the crystallography (eg crystal phase, orientation) within each segmented region. Two analysis strategies have been explored; principal component analysis (PCA) and k-means clustering. The intensity at individual (binned) pixels on the detector were used as the variables defining the multidimensional space in which each pattern in the map generates a single discrete point. PCA analysis alone did not work well but rotating factors to the VARIMAX solution did.  K-means clustering also successfully segmented the data but was computational more expensive.  The characteristic patterns produced by either VARIMAX or k-means clustering enhance weak patterns, remove pattern overlap, and allow subtle effects from polarity to be distinguished.  Combining multivariate statistical analysis (MSA) approaches with template matching to simulation libraries can significantly reduce computational demand as the number of patterns to be matched is drastically reduced.  Both template matching and MSA approaches may augment existing analysis methods but will not replace them in the majority of applications.

# 1 Introduction

Multivariate statistical analysis (MSA) tools are widely used across science disciplines including microscopy and diffraction methods. MSA methods have been developed to aide analysis of datasets that have grown increasingly large and complex. MSA aims to help extract the most useful information within the data. This is achieved by revealing hidden ('latent') variables that better and more simply describe the data, or by identifying clusters within the data that act in similar ways. This often leads to a reduction in the overall quantity of data while improving the information content or quality of the data retained. MSA has been applied to data from a range of microscopical techniques initially within the biological community, but increasingly within the materials sciences. The area where most application has been found in electron microscopy of materials is analysis of spectroscopy data obtained from the transmission electron microscopy (TEM) [1-8], though there have been notable applications in the scanning electron microscope to both energy dispersive X-ray spectroscopy (EDX) [9-11] and cathodoluminescence (CL)[12].

Electron backscatter diffraction (EBSD) is a scanning electron microscope (SEM) based technique that allows crystallographic information to be obtained from small volumes of material. The technique has been reviewed several times previously [13-18]. The technique finds widespread application in materials and earth/planetary sciences. Its huge versatility in mapping orientation [13, 19-23], crystal type [24-28], strain [29-33], dislocation content [34-41] and perfection over a wide range of length scales makes it a powerful microstructural characterization tool.

To date, the application of MSA methods to EBSD data has been surprisingly limited. Indeed the work of Brewer, Kotula, and Michael [42] appears to be the only previous publication in the area and is now ~10 years old, though a patent application by Stork and Brewer [43] describing a general hierarchical clustering method mentions potential application to EBSD data. Brewer et al demonstrated the potential for MSA in reducing the number of EBSD patterns to be indexed, and improving their quality (i.e. signal to noise ratio, and thus indexibility). They foresaw the utility of the approach in deconvolving pattern overlap (a pattern comprising information from more than one crystal) issues that occur in fine grained materials where MSA might extend the capabilities of EBSD.

In applying MSA to EBSD for microstructural analysis the potential benefits are in (i) segmenting the spatial domain into distinct similar regions (i.e. grains, sub-grains, or domains), and (ii) producing a smaller number of representative lower noise patterns associated with these regions. Implementing the first benefit should allow the morphology of the microstructure to be revealed and quantified without the requirement to index the patterns themselves. This would allow, for example, quantification of grain size and shape distributions to be undertaken. This may be advantageous in situations when the individual EBSD patterns themselves are too weak or blurred to be indexed with confidence. The possibility to assign each pixel to a single cluster with other pixels generating similar EBSD patterns directly addresses the pattern overlap issue and may provide sharper definition of grain boundaries. Most MSA schemes also provide measures of signal strength contributed from each of a set of neighbouring grains which offers a route to interpolate and better refine the boundary position independent of the EBSD grid points.

The second potential benefit of MSA analysis is the generation of a smaller number of better quality EBSD patterns. Brewer et al [42] demonstrated that these representative patterns could then be indexed using established Hough-transform based methods. This has the added benefits that signal to noise is improved and a time is saved from the reduced number of patterns to be indexed. The improved pattern quality may have real advantages when pattern quality is low due to sample preparation issues, high deformation states or beam sensitive samples, or when fine details in the patterns need to be used to resolve pseudo-symmetry issues. An important new possibility that has emerged since the work of Brewer et al is indexing via template matching to simulated pattern libraries. Here, the number of test patterns to compare against is reduced, providing a major reduction in the computational time required. It seems unlikely that MSA will replace Hough-based analysis of EBSD patterns but it may augment it by improving analysis of low quality data sets, or differentiating finer details not routinely detected through the Hough/Radon transform.

The MSA methods will be described in section 2, and the template matching analysis for pattern indexing in section 3, with the main steps illustrated using a small data set from a ferritic steel sample. Further examples will then be given in section 4 to demonstrate some benefits obtained by employing MSA in the analysis of EBSD and TKD (transmission Kikuchi diffraction) data.

# 2 Multivariate Statistical Analysis

## 2.1 Principal Component Analysis

EBSD patterns recorded in a map can be considered as a set of independent intensity measurements at each pixel (or group of binned pixels) on the detector. Perhaps the most commonly used of the MSA methods is principal component analysis (PCA) [44]. PCA aims to convert a set of observations (eg patterns in an EBSD map) of a number of variables (eg intensity at each detector pixel) into a set of values of linearly uncorrelated (latent) variables called principal components. An orthogonal transformation is used such that the first principal component has the largest possible variance from the mean i.e. it describes the greatest variability in the data. Succeeding components are then sequentially established so that they are orthogonal to the preceding components and have the highest variance possible. The resulting principal component vectors form an uncorrelated orthogonal basis set. PCA can be used to reduce the dimensionality of a problem by retaining only a subset of the most significant principal components.

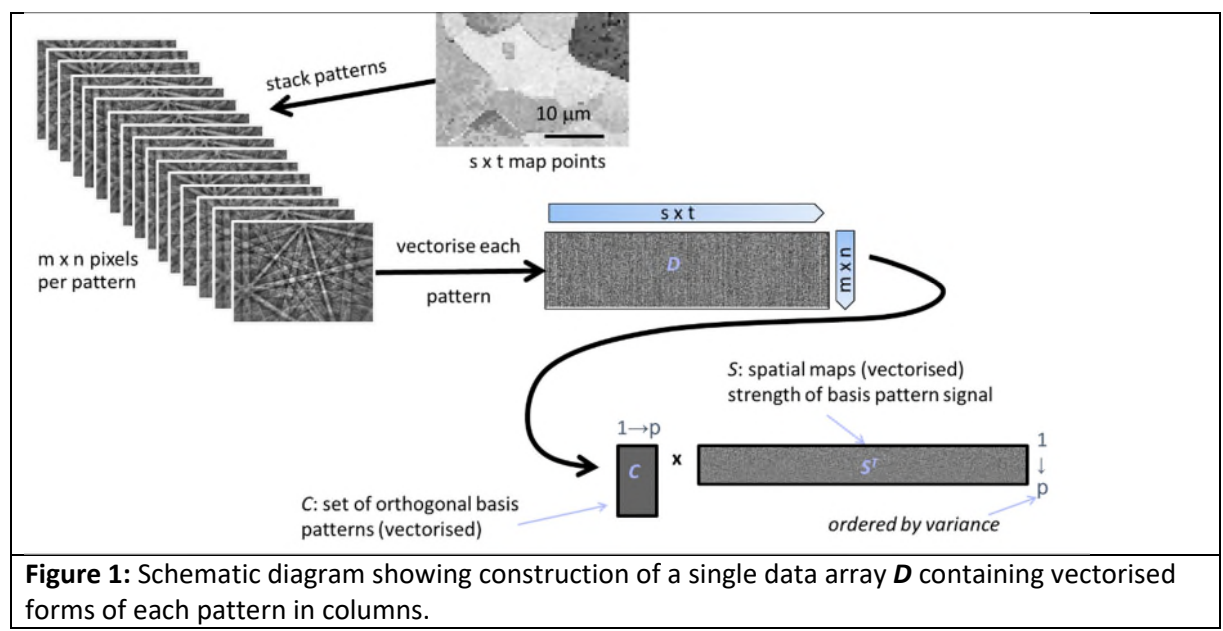

**Figure 1:** Schematic diagram showing construction of a single data array *D* containing vectorised forms of each pattern in columns.

To apply PCA to an EBSD dataset the first step is to re-write each EBSD pattern (m×n pixels) as a vector containing intensities at each effective pixel after binning on the camera. The set of patterns are then arranged into a data array **D** with columns containing intensities from an individual pattern (i.e. a single observation of the many m×n variables), and rows containing the intensity variation across the s×t points of the spatial map for a particular detector pixel (i.e. all s×t observations for a single variable). Construction of the data array is illustrated in figure 1. The data array can be large and so significant binning on the detector is advised. We then seek to decompose this data array into a set **C** of characteristic vectors describing the underlying intensity distributions on the detector (i.e. basis patterns) and **S** the variation of their strength from one observation point to the next (i.e. spatial maps, where the superscript T denotes a matrix transpose):

$$D_{[m\times n, s\times t]} = C_{[m\times n, p\leq m\times n]} S^T_{[p\leq m\times n, s\times t]} \qquad (eq\ 1)$$

There are m×n characteristic vectors, which were determined using singular value decomposition within MatLab. Data reduction can be implemented by only retaining the first p ($p \leq m \times n$) of these eigen-vectors each of which can be reformed into a characteristic EBSD pattern.

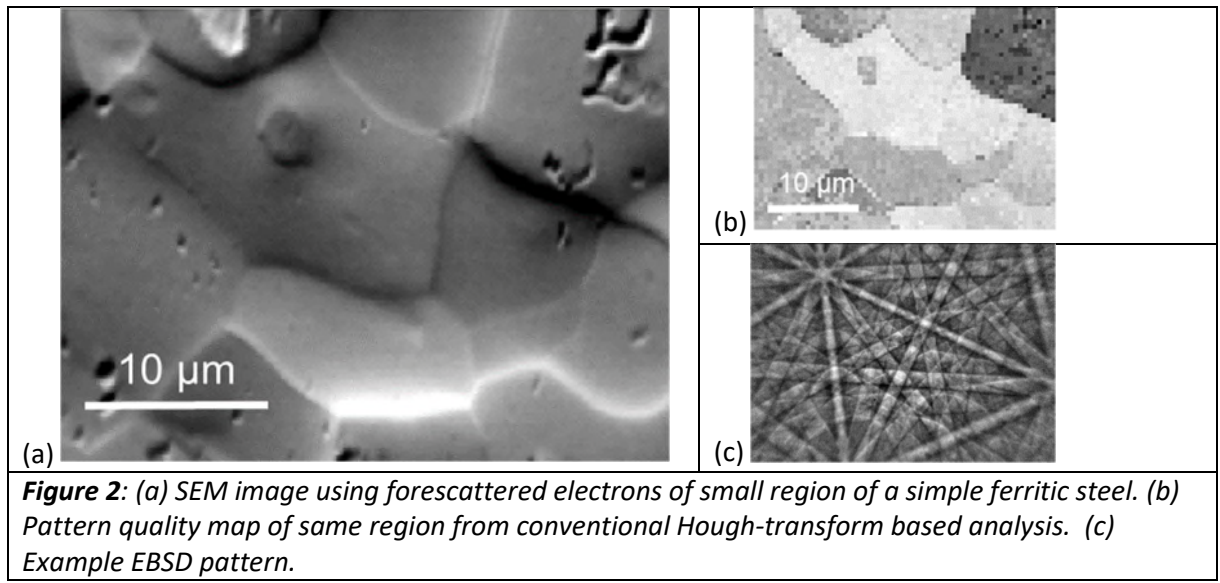

**Figure 2**: *(a) SEM image using forescattered electrons of small region of a simple ferritic steel. (b) Pattern quality map of same region from conventional Hough-transform based analysis. (c) Example EBSD pattern.*

To demonstrate the result of this decomposition we will use a small data set obtained from a ferritic steel sample using a Bruker e-Flash EBSD detector in Zeiss Merlin FEG-SEM system. The patterns were collected at a camera resolution of 800×576 pixels over a square grid of 64×48 points on the sample. The raw recorded patterns were read into MatLab, a median filter applied to remove a small number of hot pixel defects and the result further binned down to 400×288 pixels. The background intensity was then flat-fielded by applying a Gaussian kernel filter to generate a 'dynamic' background which then removed from the pattern by division. Figure 2 shows an SEM image of the mapped region along with a pattern quality map and an example EBSD pattern after background correction. Some of the basic output from the PCA analysis is given in figure 3. The left hand column shows the first 4 characteristic patterns that are found by the PCA method, along with the corresponding spatial map showing how the strength of this contribution varies across the sample in the second column. Generally, the characteristic patterns obtained through PCA have no physical meaning and correspond to a mixture of overlaid patterns from the various grains in the sample, including some EBSD patterns with inverted contrast. The corresponding maps show that the strength of signal for these characteristic patterns is generally distributed at different level across many of the grains. Of course, PCA seeks to maximise the variance in the signal variation, so the maps in the second column sequentially decrease in intensity deviations. The right hand column of figure 3 shows an attempt to segment the grain structure. The upper map shows the ratio between the highest and second highest signals contributing to EBSD patterns obtained at each point. In general, this ratio is not very high (mean below 2 and all points below 10 in this example). However, if each pixel is assigned a number corresponding to the characteristic pattern contributing most strongly to the signal, an attempt can be made to segment the grain structure. This is shown in the lower image in the right hand column of figure 3. It is obvious that direct application of PCA does not separate the data in any physically interpretable way.

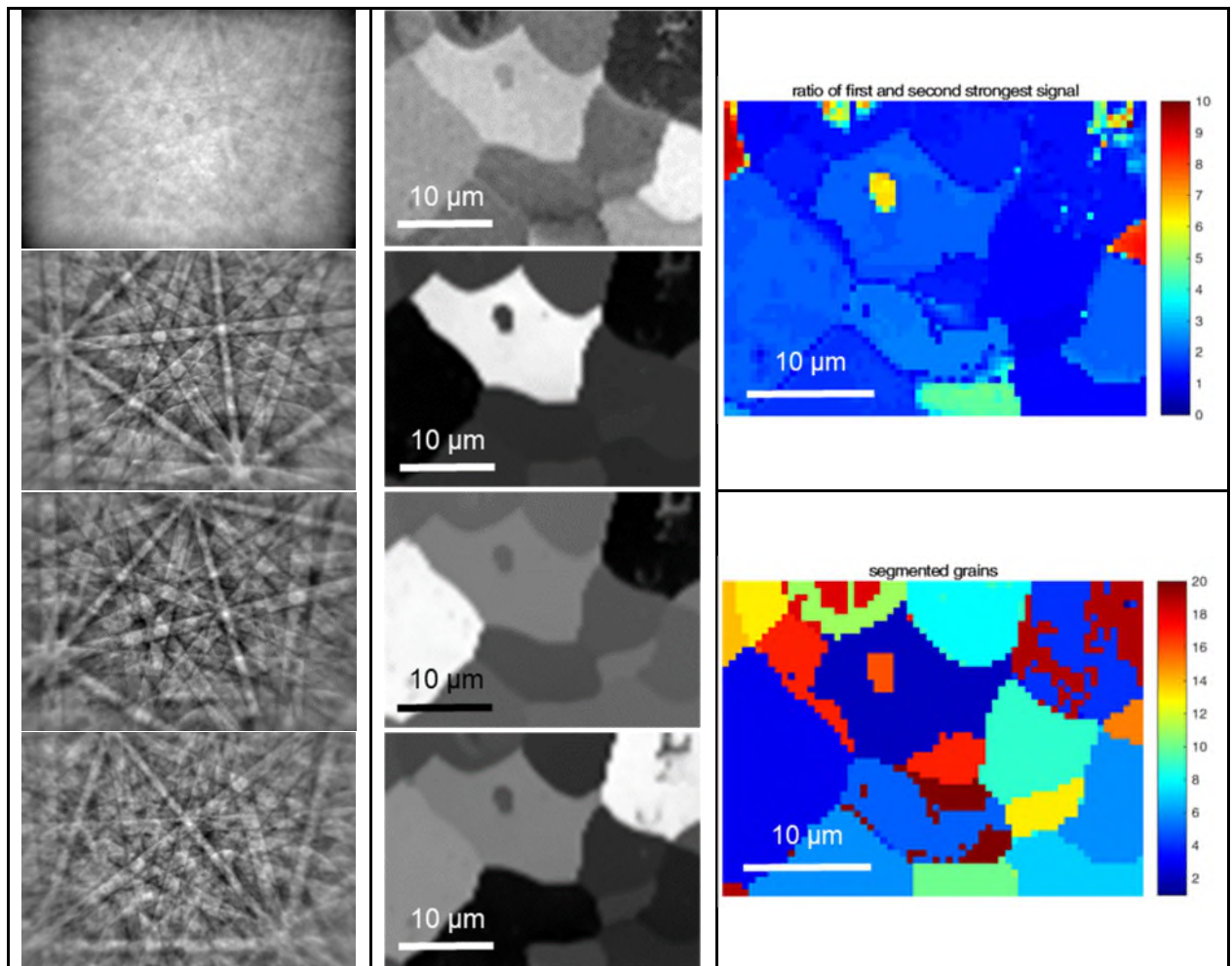

*Figure 3:* Principal component analysis of example EBSD dataset from ferritic steel. Left hand column shows the first four characteristic patterns obtained from principal component analysis, with maps showing the spatial variations in signal corresponding to these patterns given in the second column. The upper right hand image shows the ratio between the strongest and second strongest contributing components to the signal level and the lower right hand image shows the segmentation of the grain morphology into the 20 available principal components.

## 2.2 VARIMAX Component Analysis

A problem evident in figure 3 is that the choice of orthogonal basis vectors in PCA analysis does not guarantee a single spatial point in the map will be dominated by a single characteristic EBSD pattern. This issue was noted by Brewer et al [42] and is mentioned by Bonnet [1]. The VARIMAX solution [44], also used by Brewer et al, is more suitable as this seeks basis vectors such that each observation (i.e. EBSD map point) tends to be dominated by a single basis vector (i.e. characteristic EBSD pattern). Following the PCA decomposition given by eq 1, we seek a (hyper-) rotation **R** such that the sum of squared correlations between and the new basis patterns (**C R**) and their spatial variation (**S R**) is maximised.

$$\boldsymbol{D}_{[m\times n, s\times t]} = \boldsymbol{C}_{[m\times n, p\leq m\times n]} \, \boldsymbol{R}_{[p,p]} \, \boldsymbol{R}^T_{[p,p]} \, \boldsymbol{S}^T_{[p\leq m\times n, s\times t]} \qquad (eq\ 2)$$

Figure 4 shows the first 3 characteristic patterns after rotation to the VARIMAX solution; these are all high quality EBSD patterns, with each from a single diffracting crystal. Occasionally, the output from the VARIMAX rotation gives a characteristic pattern with inverted contrast. This is always linked to negative signal levels in the spatial map which are unphysical. To correct for this, the EBSD pattern contrast and signal are inverted if the mean signal for that part of the map from a particular characteristic pattern is found to be negative. The maps in the second column show the signal strength from the corresponding characteristic patterns in column one. The desired result is obtained that high signal is observed from a contiguous region of the map, and very low signal elsewhere. The ratio of strongest to second strongest signal is greatly increased (mean of ~500, and some values well in excess of $10^5$ for this example), which makes for clearer segmentation of the grain structure.

Comparing the grain segmentation in figure 4 with the pattern quality map and SEM image (figure 2) shows that some of the grains have been fragmented. An example of this is the large grain toward the centre of the map corresponding to characteristic pattern 1 in the VARIMAX solution. The other part of this grain is dominated by another characteristic pattern (number 17) which is shown at the bottom of figure 4. Characteristic patterns 1 and 17 are clearly visually very similar to a human observer (though the corresponding vectors are still orthogonal) and differ only by a small misorientation. The corresponding grain maps show the signal swapping gradually from one characteristic pattern to the other in a systematic way. There are other grains in this map that have similar partitioning into multiple characteristic patterns that are closely related in their physical origin. This demonstrates that MSA can reliably detect quite subtle changes in the input EBSD data. Once the characteristic patterns are indexed then the grain structure fragmented by the segmentation can be reassembled into grains of closely shared orientation.

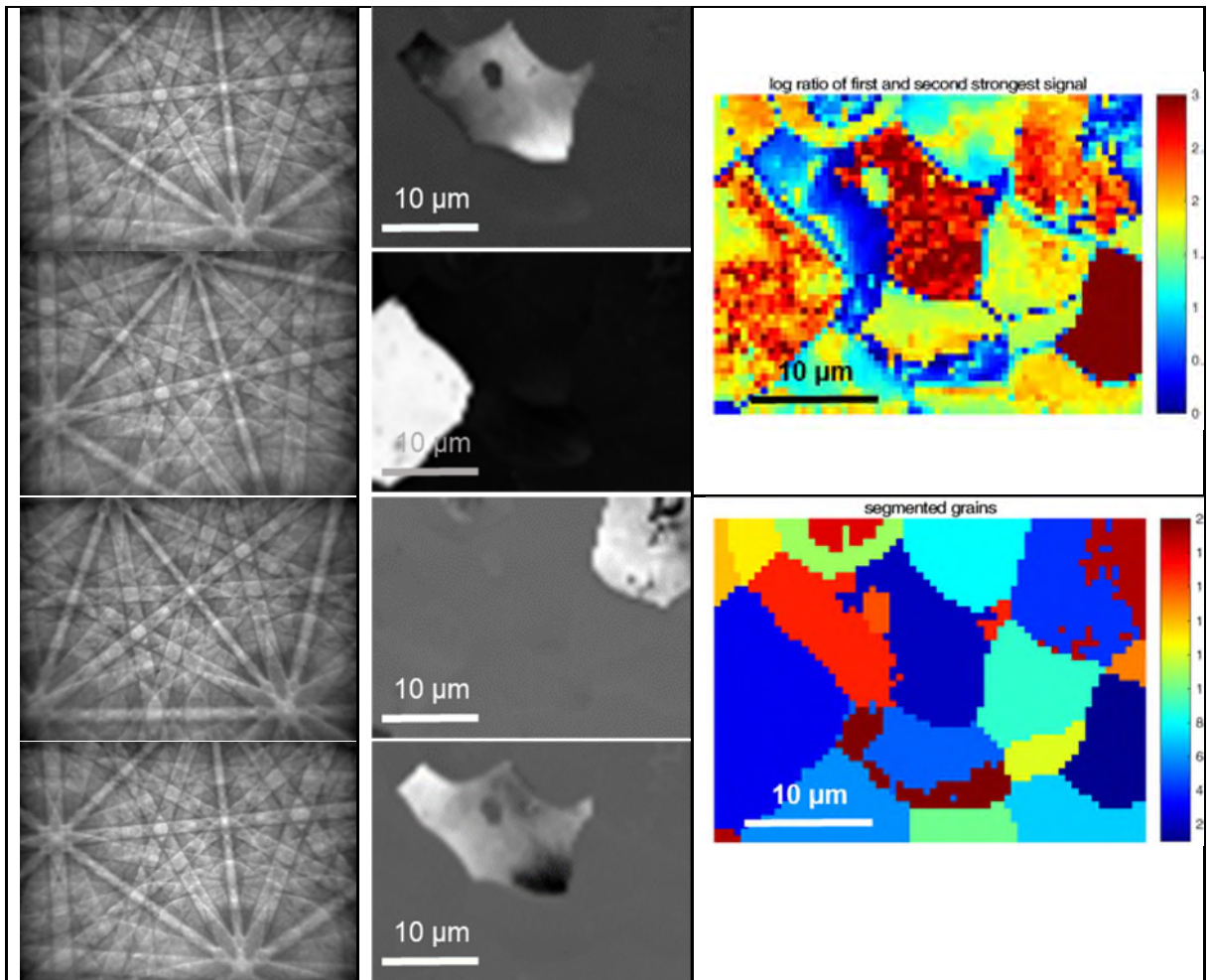

*Figure 4: Rotation of the PCA output to the VARIMAX solution for the example EBSD dataset from ferritic steel. Left hand column shows the first three characteristic patterns (plus pattern 17), with maps showing the spatial variations in signal corresponding to these patterns given in the second column. The upper right hand image shows $\log_{10}$ of the ratio between the strongest and second strongest contributing components to the signal level and the lower right hand image shows the segmentation of the grain morphology into the 20 available VARIMAX components.*

## 2.3   K-means Cluster Analysis

An alternative method of segmenting the data is through the k-means clustering approach [45]. Here, the aim is to assign each observation to one, and only one, of a finite number of partitions informed by the distance between the observations (i.e. points in the map) in a multi-dimensional space with axes representing each of the measurement variables (i.e. intensities at each pixel on the EBSD detector). The partitioning is performed to minimise the sum, over all data points, of the distances of an observation from the mean value for that partition. An iterative scheme is used for this minimisation, and the means, or centroid positions for each partition then serves as its characteristic observation (i.e. characteristic EBSD pattern). The iterative scheme for establishing the clusters often finds local minima from which reassigning an observation to another cluster will increase the point to centroid distance. To reduce the sensitivity to the initial choice of cluster centroid positions, the process can be re-seeded from several different initial configurations (here 50 were used) and then the lowest of these local minima used as the final output.

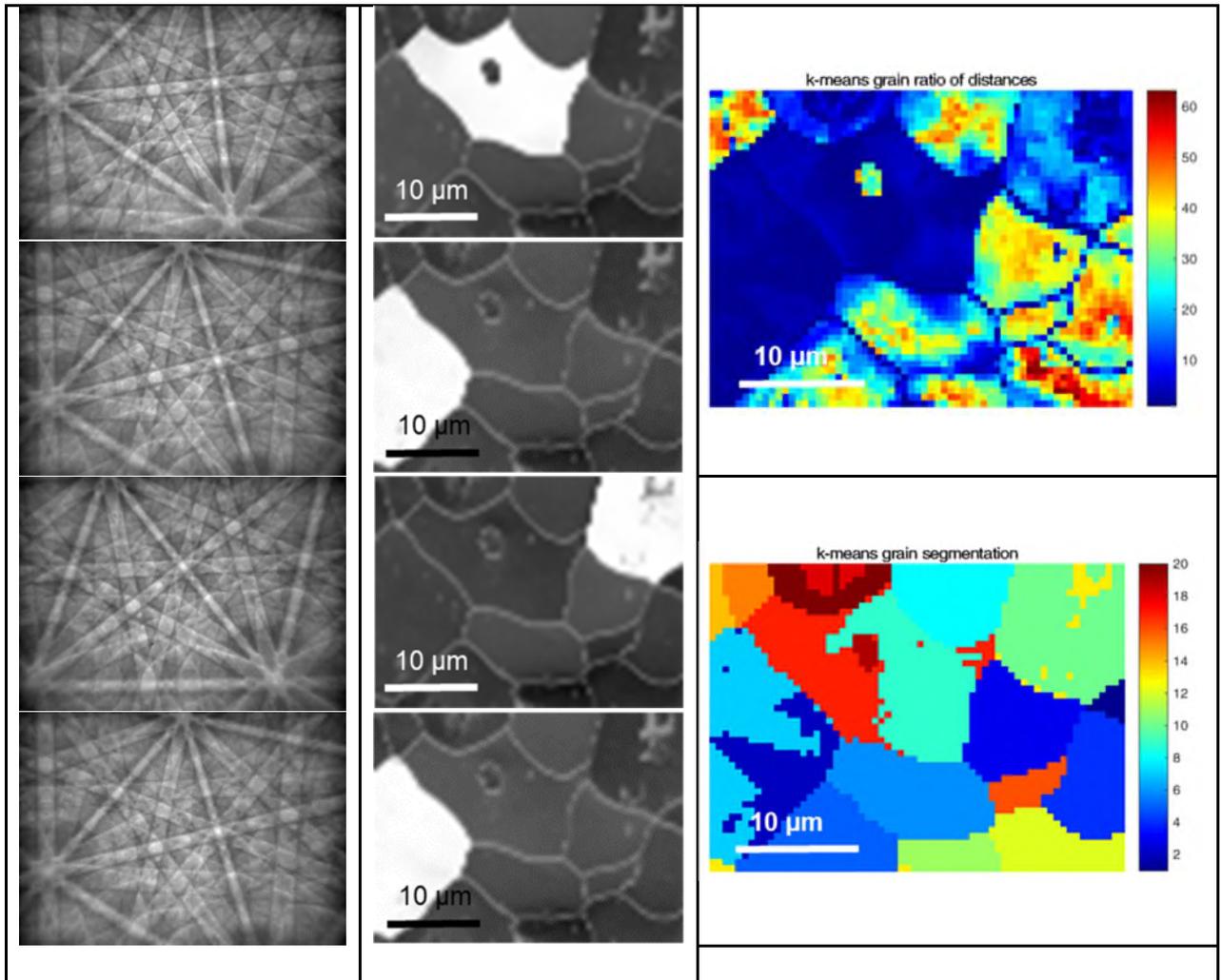

*Figure 5: k-means cluster analysis for the example EBSD dataset from ferritic steel. Left hand column shows the patterns obtained from four of the centroids of clusters, with maps showing the distance of the individual pattern intensities from the centroid in the second column (white→at centroid, black→far from centroid). The upper right hand image shows ratio between distance from the second nearest and nearest centroid and the lower right hand image shows the segmentation of the grain morphology into 20 domains.*

The steel dataset was also used to illustrate the output from k-means clustering analysis. The left hand column of figure 5 shows the EBSD patterns corresponding to the centroids of four of the clusters, along with corresponding maps (second column) which plot the proximity of EBSD pattern intensities to the centroid as a function of position on the sample. The centroid positions do represent good quality EBSD patterns from a single grain, and the distance maps show that the grain morphologies are well captured too. The ratio of the distance metrics for the second nearest to the nearest centroid gives a measure of how confidently each point can be assigned to a particular grain, and this is shown at the top of the right hand column. At the bottom of the right hand column is the segmentation of the grain structure achieved via the k-means clustering route, on the assumption that 20 grains should be found (note the implications for the chosen number of grains is discussed later in Sections 4.1 and 4.2). As with the VARIMAX approach some of the grains that can be seen in the original SEM image have been fragmented into multiple domains. The large grain to the left hand side of the image (rows 2 and 4 of figure 5) give an example of this, where it can be seen that the patterns formed for the two clusters are visually very similar.  Indexing of these patterns would result in similar orientations and a re-joining of the data into a single grain.  For the k-means clustering approach there is no restriction for the solutions (cluster centroids) to be orthogonal.  This reflects the physical situation where no condition of orthogonality exists between observed patterns and is a significant advantage compared to the PCA and VARIMAX approaches.

# 3 Pattern Matching to Simulation Libraries

Rauch and co-workers (eg [46-49]) developed a template matching approach to index spot diffraction patterns obtained in TEM.  A library of possible diffraction patterns is calculated for crystal orientation dispersed over the range of possible orientations and compared using correlation metrics to the experimentally observed patterns.  More recently, this approach has been adapted to analysis of EBSD patterns [50-53] with good success.  The task is more demanding for EBSD than for spot diffraction patterns as dynamical theory has to be employed for EBSD while for the TEM case, kinematic theory suffices especially if precession is employed for data collection.  Dynamical diffraction theory based simulations of EBSD patterns pioneered by Winkelmann [54-56] and more recently by Callahan & De Graef [57] have been available for a decade but in recent years are becoming much more frequently exploited in advanced analysis of EBSD data [50-52, 58].  One consequence of the greater computational demand for simulating EBSD patterns is that calculating each individual EBSD with the full theory is prohibited and instead a 'master' pattern that covers all crystallographically distinct directions is calculated. From this, interpolation is used to construct individual patterns at lower computational cost.

In this work the Bruker Dynamics software was used to simulate the master EBSD patterns onto a stereographic projection.  The master pattern was read into MatLab and bi-cubic interpolation of the intensities of the stereographic projection was then used to populate the individual EBSD patterns on a regular square grid on the gnomonic projection.  A random distribution of crystal orientations was generated using the MTex toolbox [59] within MatLab.  MTex was also used to generate pole figures, inverse pole figures, and inverse pole figure maps presented in this paper.

Figure 6a shows, in stereographic projection, the master pattern calculated for Fe using the Bruker Dynamics software.  An example target experimental pattern to be matched is shown in figure 6b (taken from fig 4), along with the best four matches (figure 6c) from the simulated template library which consisted of 500,000 templates of size 100x72 pixels.  The pattern centre was assumed to be fix at 0.477 of the detector width from the left, 0.714 of the detector height from the bottom and a sample to screen distance of 0.813 of the detector height.  The normalised correlation coefficient was used for ranking the matches, equivalent to the largest normalized inner product of the vectorised forms of the patterns generated during the MSA analysis described in section 2.  Figure 6d shows the correlation coefficient on an (100) pole figure indicating the simulated crystal orientation which shows clear and well localised peaks which allow good confidence in identifying the correct crystal orientation from the template library.  The correlation coefficient is shown in figure 6e as a function of the disorientation angle between the crystal orientation corresponding to the particular template and that giving the best match.  The cross-correlation coefficient falls to below half its peak value for disorientations beyond less than ~2°.  The width of this peak places limitations on the minimum size of the template library since it must be constructed so that at least one template is sufficiently close to the target value for the peak to be detected.

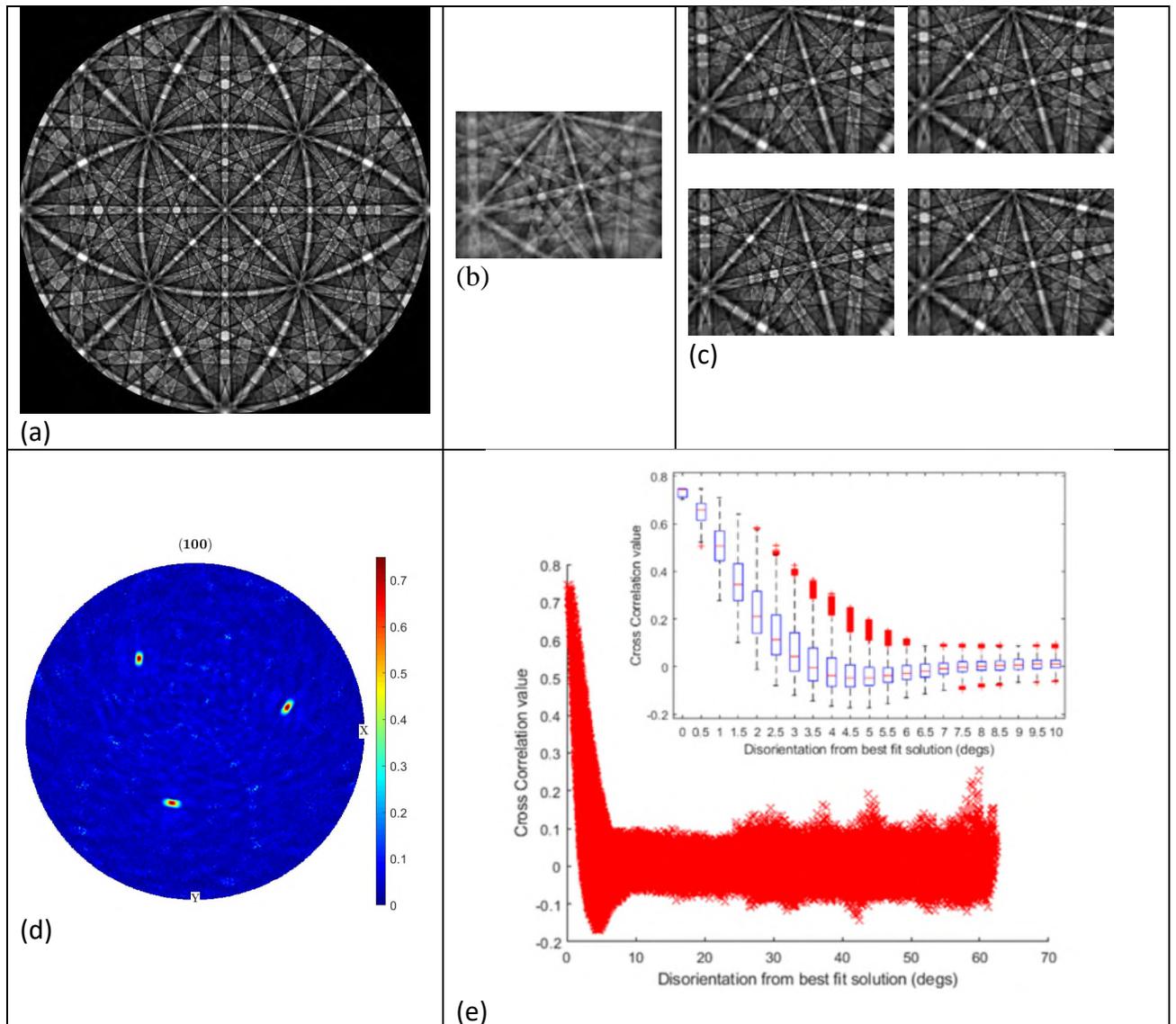

***Figure 6:*** *(a) Master pattern on stereographic projection calculated using Bruker Dynamics software, (b) target experimental EBSD pattern from VARIMAX analysis (c) best four matches (in order) from template library, (d) correlation coefficients for templates patterns showing narrow peak of good matching on (100) pole figure, and (e) the disorientation from the best fit solution.*

## Areas of Potential Application

### 4.1  Segmentation of Grain Structure

As noted in section 2, for both PCA and k-means clustering approaches there is a need to set the number of basis components or clusters to separate the data into. This has the potential to strongly influence any measures of the microstructures produced. This is illustrated using an EBSD dataset obtained from a commercially pure alpha Zirconium sample (CP-Zr) using a Bruker eFlash EBSD detector on a Zeiss EVO SEM and consisting of ~0.9M patterns each binned to 50x50 pixels size for the VARIMAX analysis. A pattern quality map from conventional Hough-based indexing is given at the centre of figure 7. Figure 7 also shows how the perceived grain size varies with the number of basis components retained in the VARIMAX analysis. The mean (by area) equivalent circle diameter $D_{ecd}$ varies significantly and non-linearly with the number $p$ of components retained in the analysis. For small $p$ where there are too few components retained to account for the grains individually $D_{ecd}$ increases markedly as $p$ is decreased, $D_{ecd}$ is then relatively constant over the domain $150 \leq p \leq 200$, while at higher $p$ a gradual decrease with increasing p is seen. Identifying an appropriate value for $p$ is obviously the key to any quantitative analysis.

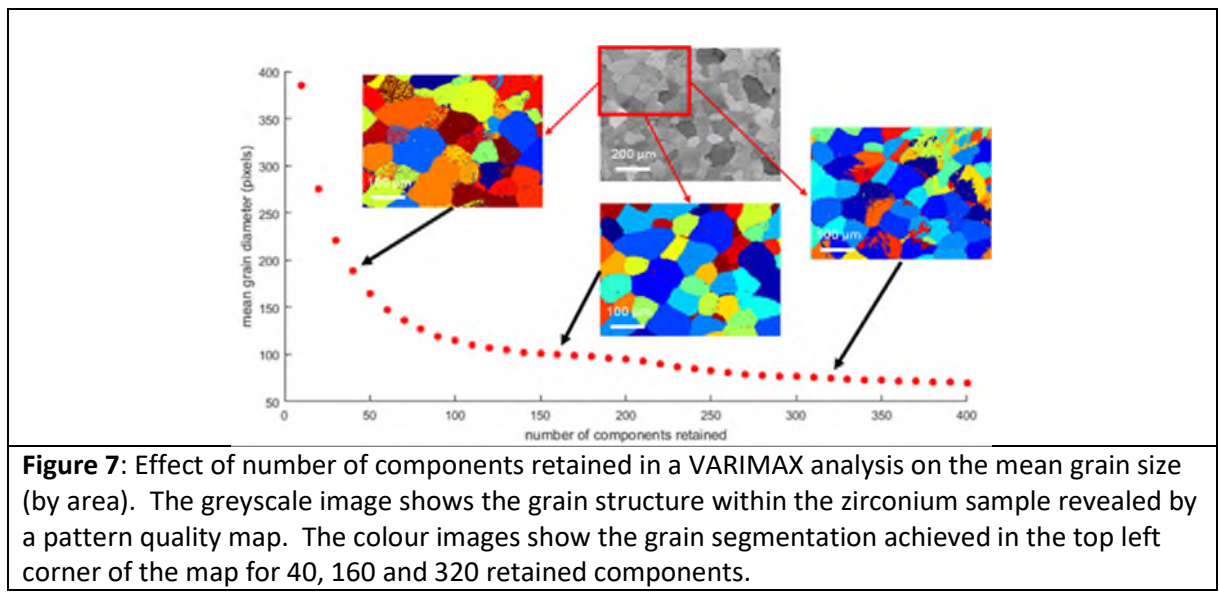

**Figure 7**: Effect of number of components retained in a VARIMAX analysis on the mean grain size (by area). The greyscale image shows the grain structure within the zirconium sample revealed by a pattern quality map. The colour images show the grain segmentation achieved in the top left corner of the map for 40, 160 and 320 retained components.

The so-called scree plot is often used in MSA to assess an appropriate number of components to retain, and is a plot of the fraction of the variance in the original data as a function of the component ranking. Typically, a rather subjective process of identifying a 'knee' or bend in this plot where the gradient changes significantly is employed. For EBSD datasets we often find the scree plot exhibits a rather smooth decrease leading to a large range of uncertainty in setting the number of components to retain. Figure 8a shows an example scree plot for the CP-Zr dataset.

If the segmentation is performed well then the strength of signal from the strongest component should dominate the pattern at that point in the map. Figure 8a shows that the ratio of the second strongest to the strongest signal averaged over all points in the map shows a distinct minimum at ~160 retained components. This corresponds to the plateau in figure 7 where mean grain size does not vary strongly with the number of retained components, and where visual inspection shows the segmentation to be reasonable.

When the number of retained components is too high or too low there are many more single isolated pixels that are wrongly assigned and break up what should be a contiguous grain. Such isolated points can be identified and corrected using a median filter. Simply counting the number of pixels in the map that would be changed by applying a median filter thus gives a useful metric. This is shown in figure 8b with a clear minimum at ~170 retained components. Figure 7 shows that if too many components are retained there is a tendency for what should be a single grain to be split into multiple separate components, which are often fragmented into several non-contiguous regions. A further metric that can be used to select the number of components is thus the average number of spatially unconnected regions in the map assigned to each component. This is demonstrated in Figure 8c where a minimum number of unconnected regions per component is found when ~180 components are retained.

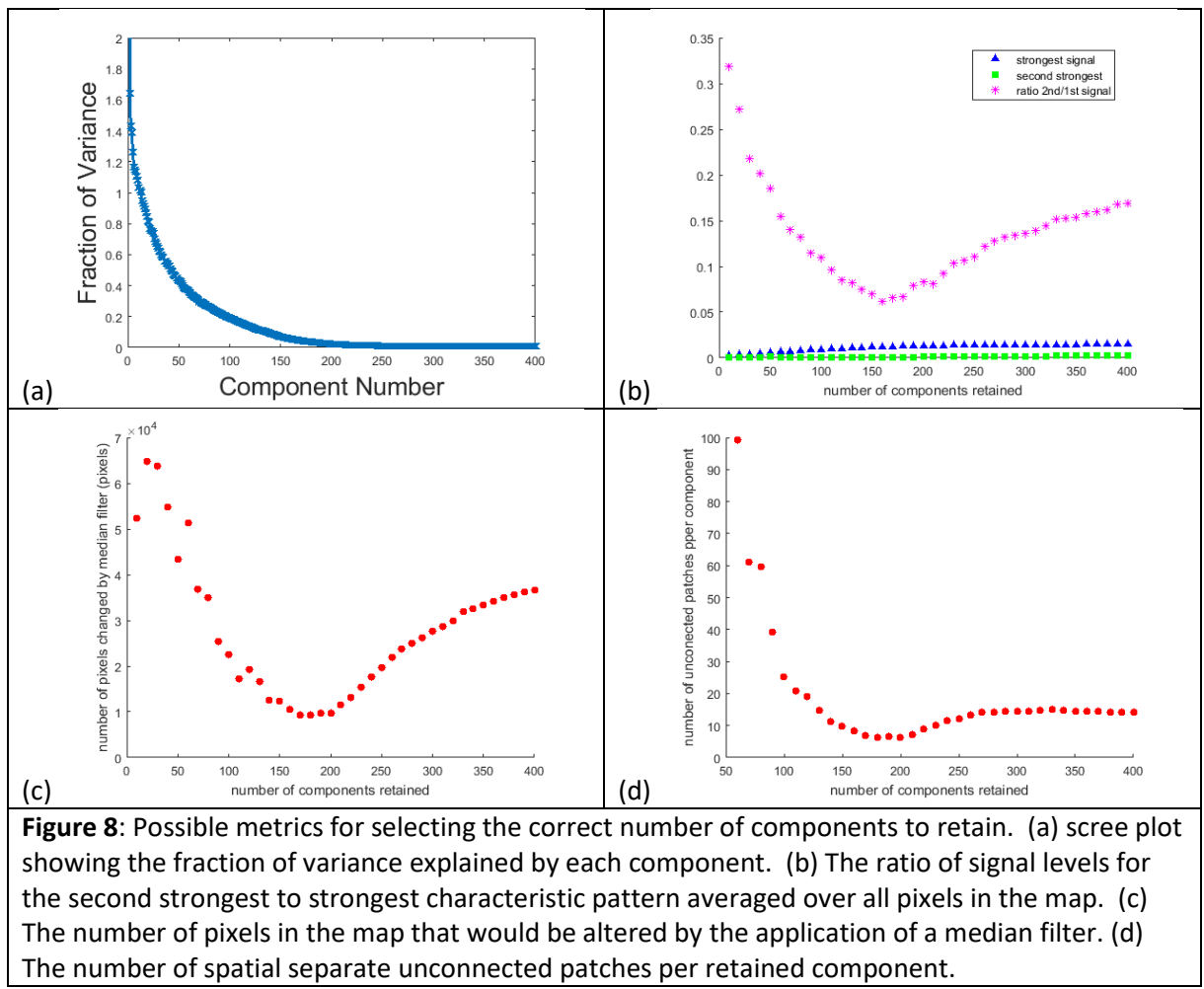

**Figure 8**: Possible metrics for selecting the correct number of components to retain. (a) scree plot showing the fraction of variance explained by each component. (b) The ratio of signal levels for the second strongest to strongest characteristic pattern averaged over all pixels in the map. (c) The number of pixels in the map that would be altered by the application of a median filter. (d) The number of spatial separate unconnected patches per retained component.

## 4.2 Time Saving for Template Matching

The template matching approach to indexing EBSD patterns described in section 3 is receiving significant interest at the current time. One of the issues that may limit the take-up of the method is the significant computational time required to run the analysis. The time required to compute the template library is of course highly dependent on the size, in pixels, of the patterns themselves and the number of them within the library. Higher angular resolution in the output orientations is strongly influenced by the number of patterns in the library as the discrete angular sampling of orientation space is often the main limiting factor. Pattern size has a much less marked effect than library size, with Ram et al [60] showing similar angular resolution for patterns of 60x60 down to 25x25 pixels.

The time taken for pattern matching itself is affected by the size and number of templates in the library but also scales linearly with the number of experimental patterns in the map. Using MSA to segment the map into grains and generate a single characteristic pattern for each grain provides an opportunity to very significantly reduce the time taken for the pattern indexing with the proviso that a single orientation per segmented grain is an acceptable quantification of the microstructure. For the small map from the steel sample used in section 2 there were 3024 points in the initial map, but this can be segmented into slightly fewer than 20 grains giving a ~1/150 reduction in the number of indexing operations to be undertaken, while for the Zr map in section 4.1, the $0.9 \times 10^6$ points is reduced by an even larger factor of ~1/5000 to ~170 grains.

Importantly, once the pattern indexing stage has been completed any fragmentation of grains due to there being too many components or clusters imposed on the data is removed again. Figure 9 illustrates this using the small map from the steel sample which was segmented into 22 regions using k-means clustering. This is (deliberately) rather more clusters than real grains present so some features appear fragmented. However, the characteristic patterns for clusters within the same grain are visually similar and so after using the template matching to extract orientations the fragments are brought back together again as can be seen in the IPF maps.

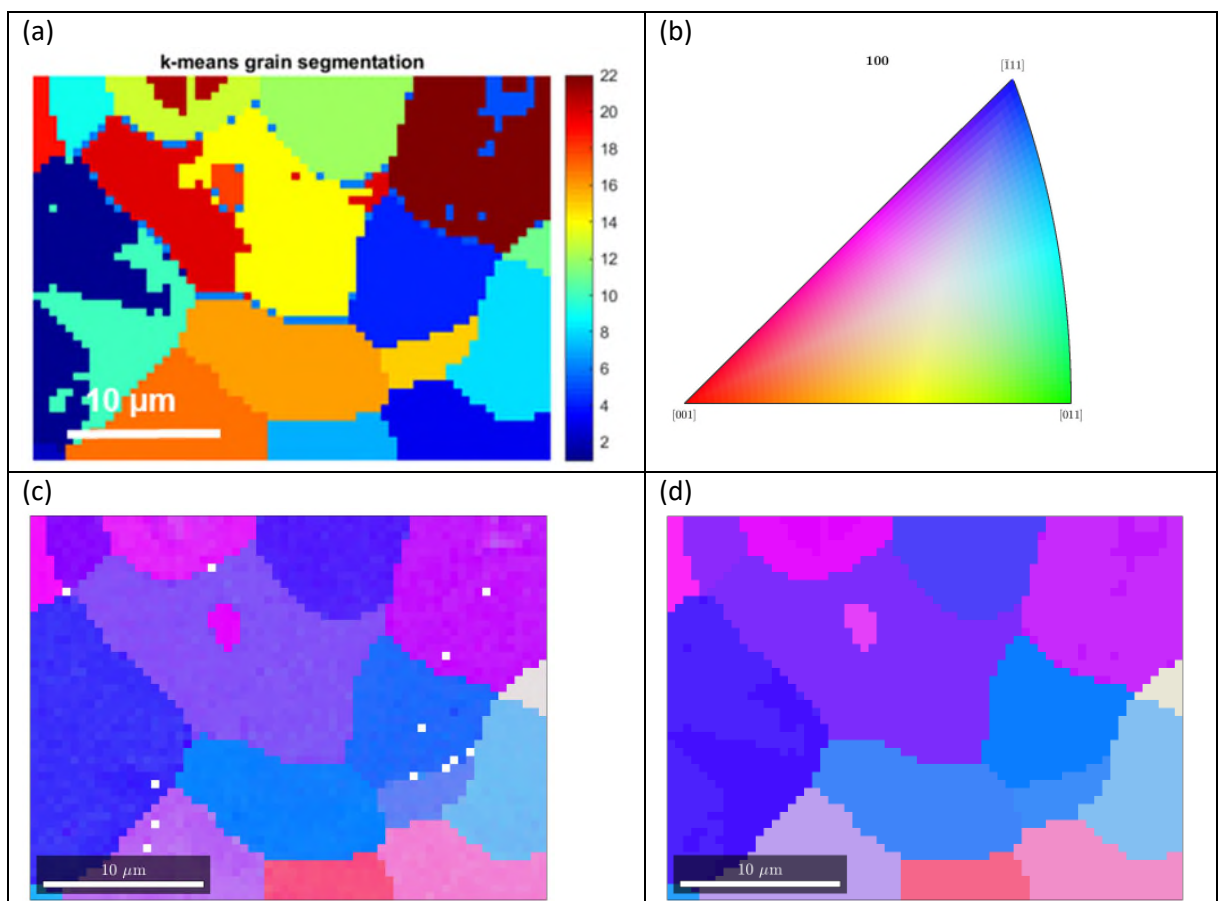

**Figure 9**: Ferritic steel sample. (a) Spatial map of k-means cluster analysis (k=22). (b) Colour key for IPF maps show crystal directions along sample surface normal obtained from (c) conventional Hough-based indexing and (d) template matching of characteristic patterns.

Running these calculations on a relatively modest PC with intel i7-3930K CPU (at 3.2GHz) and 32Gbytes of RAM running Matlab2018a under Windows 7 Enterprise achieved the following computational times. For the steel data set the PCA segmentation followed by VARIMAX analysis to

20 spatial domains and characteristic patterns at 200x144 pixels took 3 mins, while pattern matching all patterns took ~17 mins, with a projected time of over ~40 hours for the full 3024 patterns in the whole map. An additional overhead of ~138 s was taken to construct a template library containing $0.5 \times 10^6$ simulated patterns which is required in either case.

## 4.3 Improving Pattern Quality

This work was initially motivated by TKD datasets obtained from nanostructured SiC composite samples in which the poor pattern quality and low indexing rate was limiting the ability to successfully quantify the microstructure. This is illustrated in Figure 10 which shows a small dataset obtained using off-axis TKD. The low pattern quality and gnomonic distortion of the TKD patterns limit the successful indexing from conventional Hough-based analysis. This results in a map that is too sparsely populated for the microstructure to be reasonably described. Analysis of the dataset using the VARIMAX approach allows the microstructure to be segmented directly and reveals the nanostructured grains and their aspect ratio and alignment with the composite. The characteristic pattern also show significant enhancement compared to the initial patterns obtained at individual points. In some ways this analysis takes the Neighbour Pattern Averaging and Re-indexing (NPAR) approach introduced by Wright et al [61] to a limiting case in which regions generating similar patterns are intelligently segmented and the characteristic pattern generated using data from all the points within the region.

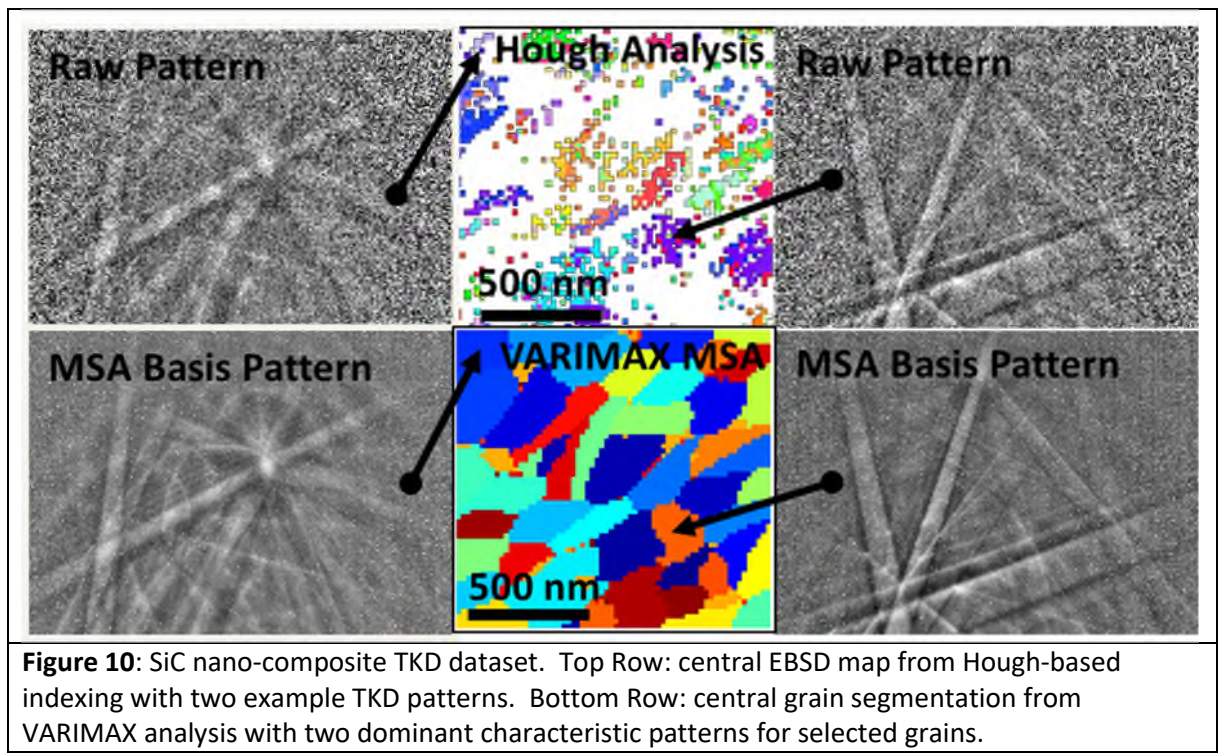

**Figure 10**: SiC nano-composite TKD dataset. Top Row: central EBSD map from Hough-based indexing with two example TKD patterns. Bottom Row: central grain segmentation from VARIMAX analysis with two dominant characteristic patterns for selected grains.

## 4.4 Extraction of fine detail

One further example application is in segmenting datasets based on rather subtle differences between EBSD patterns that are not generally detected in conventional analysis. Polarity of non-centrosymmetric crystals can lead to relatively small changes in intensity profiles across Kikuchi bands in EBSD patterns as has been demonstrated by the elegant pattern matching approach of Winkelmann and Nolze [53]. Figure 11 shows the segmentation using PCA followed by rotation to the VARIMAX solution (section 2.2) of an EBSD map, previously reported by Naresh-Kumar et al [62], of a single crystal GaP film grown on (001) Si [63]. The analysis was restricted to only two component EBSD patterns since GaP has the cubic zincblende structure which exhibits two possible crystal polarities along the [001] growth direction. The two components each dominate in close to 50% of the film as expected (figure 11b). The two characteristic patterns are shown in Figure 11c and d. Visual comparison demonstrates that the patterns appear near identical, however, plotting the difference between the intensities in the two pattern reveals that systematic variations are present (Figure 11e). Intensity profiles (figure 11f) were taken across selected Kikuchi bands and averaged within boxes shown in figure 11c and show the subtle differences between the patterns. The intensity traces across the two polar inclined {111} planes (A and B in figure 11f) show bright-dark asymmetry across the band profiles in opposite senses between the two bands. For the non-polar {002} band (figure 11f profile C) the intensity profiles for the two patterns are essentially indistinguishable.

# 5 Summary and Conclusions

This paper has explored the application of principal component analysis and k-means clustering directly to intensities measured at individual pixels in an EBSD detector as a function of position on the sample. PCA analysis did not satisfactorily segment the microstructure, however, rotating to the VARIMAX basis vectors provided a good segmentation as this seeks a spatial domain with a 'simple' structure in which the signal at any point is dominated by a single corresponding characteristic or basis pattern. K-means clustering also successfully segmented the microstructures investigated though typically was slower to compute. The characteristic or basis patterns produced were generally of much lower noise than the individual patterns initially obtained in the map affording some advantage in indexing especially for poor quality datasets from difficult samples. We also demonstrate that subtle effects such as those from polarity in non-centrosymmetric crystals can be readily detected using MSA approaches and polarity domains readily revealed in the EBSD maps. The number of characteristic patterns can be drastically reduced compared to the number in the original map and this advantage is transferred directly as a reduction in the number of patterns to be indexed. This is of particular significance for the application of template matching approaches to indexing that we have successfully employed in conjunction with MSA approaches.

Both template matching and MSA approaches are significant and valuable extensions of EBSD analysis methods allowing application to difficult samples and/or rather subtle effects. However, they should be considered as augmenting existing analysis methods which they are unlikely to replace in the majority of applications.

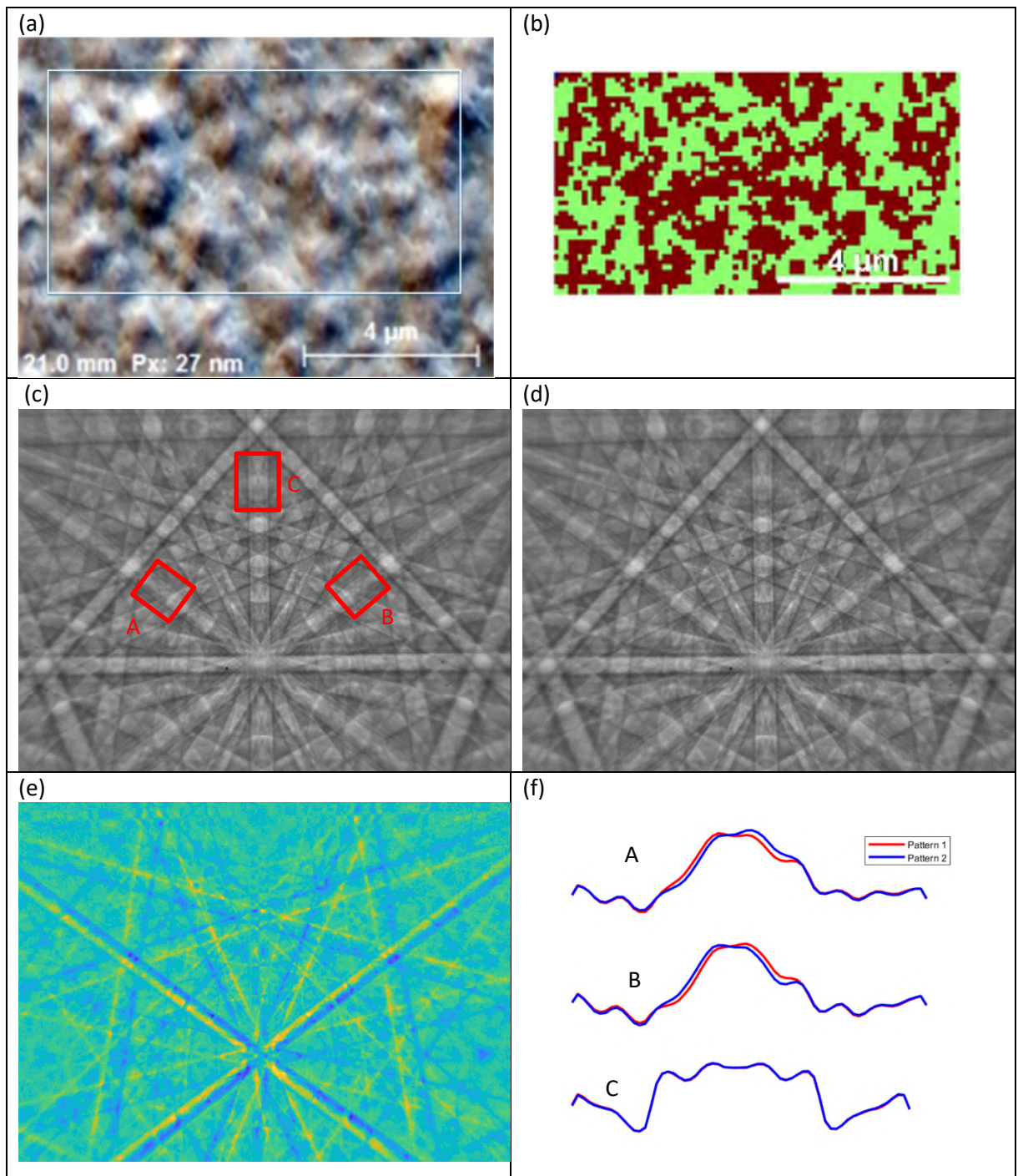

**Figure 11**: Segmentation using the VARIMAX approach of polarity domains in a GaP film grown on Si. (a) SEM image of the sample surface and area mapped, (b) successful segmentation into polarity domains, (c) and (d) show the two characteristic patterns, (e) shows a simple subtraction of the two characteristic pattern intensities to make the differences clearer, and (f) compares intensity profiles extracted across selected bands in the two patterns.

# 6 Acknowledgements

We acknowledge activities on several grants funded by the Engineering and Physical Sciences Research Council in the UK as contributing to this work: the HexMat programme grant (EP/K034332/1 –DMC/RK/AJW), nanoscale characterisation of nitride films

(EP/J016098/1 – AV-C/AJW), non-proportional deformation of steels (EP/I021043/2 – DMC/AJW), SiC nuclear fuel cladding (EP/N017110/1 - YZ).  We thank Aimo Winkelmann for helpful discussions on use of the Bruker Dynamics simulation tool.  Data used in this paper is openly available on Zenodo (http://doi.org/10.5281/zenodo.1288431).

# 7   Author Contributions

AJW developed and coded the MSA analysis approaches.  Data collection and analysis were conducted by DMC for the steel, RK for the zirconium, YZ for the SiC, and AV-C for the GaP.  The initial manuscript was drafted by AJW, with all authors contributing to the final version.